\documentclass[11pt]{article}
\usepackage{amsfonts, epsfig}
\addtocounter{page}{0}
\begin{document}

\begin{center}
{\bf The Differential Geometry and Physical Basis for the Applications of
Feynman Diagrams\footnote{ This paper appears in published form in {\it The
Notices of the AMS}. {\bf 53} 744 (2006) and the Chinese translation in {\it
Mathematical Advance in Translation}, Chinese Acad. of Sciences, {\bf 2} 97, (2009).}}
\\

\vskip .3in

Samuel L. Marateck\\

Courant Institute of Mathematical Sciences\\
New York University\\
New York, N.Y. 10012\\
email: marateck@cs.nyu.edu\\
\end{center}

\vskip .3in

\begin{center}
{\bf Abstract}
\end{center}

\begin{quotation}
\noindent
This paper recalls the development of gauge theory culminating in Yang-Mills
theory, and the application of differential geometry including connections on
fiber bundles to field theory. Finally, we see how the preceding is used to
explain the Feynman diagrams appearing on the Feynman postage stamp released in
May 2005. Version 2 included the Feynman diagrams, Version 3 corrected typos
and Version 4 included an appendix for the derivation of the Yang-Mills
transformation and field strength. Version 5 indicates that the article has
been published in the Notices of the AMS and in July 2009 appears in Chinese
translation in a journal of the Chinese Academy of Sciences. It also corrects
some typos and adds to the appendix a hueristic derivation of the Yang-Mills
field strength.

\end{quotation}

\noindent
On May 11, the late Richard Feynman's birthday, a stamp was dedicated to
Feynman at the Post Office in Far Rockaway, New York City,
 Feynman's boyhood home. (At the same time, the United States Postal Service
 issued three other stamps honoring the scientists Josiah Willard Gibbs and
 Barbara McClintock, and the mathematician John von Neumann.)\\

\noindent
The design of the stamp tells a wonderful story. The Feynman diagrams on it
show how Feynman's work originally applicable to QED, for which he won the
Nobel prize, was then later used to elucidate the electroweak force.  The
design is meaningful to both mathematicians and physicists. For mathematicians,
it demonstrates the application of differential geometry; for physicists, it
depicts the verification of QED, the application of the Yang-Mills equations
and the establishment and experimental verification of the electroweak force,
the first step in the creation of the standard model. The physicists used gauge
theory to achieve this and were for the most part unaware of the developments
in differential geometry. Similarly mathematicians developed fiber-bundle
theory without knowing that it could be applied to physics. We should, however,
remember that in general relativity, Einstein introduced geometry into
physics. And as we will relate below, Weyl did so for electromagnetism.
General relativity sparked mathematicians interest in parallel transport,
eventually leading to the development of fiber-bundles in differential
geometry. After physicists achieved success using gauge theory, mathematicians
applied it to differential geometry. The story begins with Maxwell's
equations. In this story the vector potential {\bf A} goes from being a
mathematical construct used to facilitate problem solution in electromagnetism
to taking center stage by causing the shift in the interference pattern in the
Aharonov-Bohm solenoid effect. As the generalized four-vector $A_\mu$, it
becomes the gauge field that mediates the electromagnetic interaction, and the
electroweak and strong interactions in the standard model of physics -- $A_\mu$
is understood as the connection on fiber-bundles in differential geometry. The
modern reader would be unaccustomed to the form in which Maxwell equations
first appeared. They are easily recognizable when expressed using vector
analysis in the Heaviside-Gibbs formulation.\\

\noindent
{\bf Maxwell's Equations}

\noindent
The equations used to establish Maxwell's equations {\it in vacuo} expressed in
Heaviside-Lorentz rationalized units are:\\

\noindent
(1) $\nabla \cdot \vec {\bf E}  = \rho$ (Gauss's law)\\
(2) $\nabla \cdot \vec {\bf B}= 0$ (No magnetic monpoles)\\
(3) $\nabla \times \vec {\bf B}= \vec {\bf J}$ (Ampere's law)\\
(4) $\nabla \times \vec {\bf E}= - \partial \vec {\bf B}/ \partial {t}$
((Faraday's and Lenz's law)\\

\noindent
where $\vec {\bf E}$ and $\vec {\bf B}$ are respectively the electric and
magnetic fields; $\rho$ and $\vec {\bf J}$ are the charge density and electric
current.  The continuity equation which dictates the conservation of charge:\\

\noindent
(5) $\nabla \cdot \vec {\bf J} + \partial {\rho}/ \partial {t} = 0 $\\

\noindent
indicates that Maxwell's equations describe a local theory since you cannot
destroy a charge locally and recreate it at a distant point
instantaneously. The concept that the theory should be local is the
corner-stone of the gauge theory used in quantum field theory, resulting in
the Yang-Mills theory, the basis of the standard model.

\noindent
Maxwell realized that since: \\

\noindent
(6) $\nabla \cdot \nabla \times \vec {\bf B}= 0$\\

\noindent
equation (3) is inconsistent with (5), he altered (3) to read\\

\noindent
(3') $\nabla \times \vec {\bf B}= \vec {\bf J} + \partial \vec {\bf E}/ \partial {t} $\\

\noindent
Thus a local conservation law mandated the addition of the $\partial \vec {\bf
E}/ \partial {t} $ term. Although equations (1), (2), (3') and (4) are
collectively known as Maxwell's equations, Maxwell himself was only responsible
for (3').\\

\noindent
Maxwell calculated the speed of a wave propagated by the final set of
equations, and found its velocity very close to the speed of light. He thus
hypothesized that light was an electromagnetic wave.  Since the curl of a
vector cannot be calculated in two-dimensions, Maxwell's equations indicate
that light, as we know it, cannot exist in a two-dimensional world. This is the
first clue that electromagnetism, is bound up with geometry. In fact equation
(6) is the vector analysis equivalent of the differential geometry result
stating that if {\bf $\beta$} is a p-form, and {\bf d$\beta$} is its exterior
derivative, then {\bf d(d$\beta$)}, or ${\bf d^2 \beta}$ = 0.\\

\noindent
Unlike the laws of newtonian mechanics, Maxwell's equations carry over to
relativistic frames. The non-homogeneous equations, (1) and (3'), become\\

\noindent
(7) $\partial_{\mu} F^{\mu \nu} = J^ {\nu}$\\

\noindent
while the homogeneous equations, (2) and (4) become\\

\noindent
(8) $\epsilon ^{\alpha \beta \gamma \delta} \partial_{\beta} F_{\gamma \delta}
    = 0$\\

\noindent 
where $\epsilon ^{\alpha \beta \gamma \delta}$ is the Levi-Civita symbol,
$F^{\mu \nu} = \partial^{\mu}A^{\nu} - \partial^{\nu}A^{\mu}$, and $A^0$ is the
scalar potential and $A^i$'s ($i=$ 1, 2, 3) the components of the vector
potential $\vec {\bf A}$.  Note that both equations (7) and (8) are manifestly
covariant. Yang\footnote{Yang, C.N., (1980), {\it Physics Today}, 6 42}
remarked that equation (8) is related to the geometrical theorem that the
boundary of a region has no boundary. In a later section, we will show that
equation (8) is due to the principle $d^2\omega = 0$, where {\bf $\omega$} is a
p-form. Yang's geometrical explanation can be understood in differential
geometry terms using the generalized Stoke's theorem: $\int_M d\omega =
\int_{\partial M} \omega$, where $\omega$ is an n-form and $M$ is an $n+1$
dimension oriented manifold with boundary $\partial M$. For the purposes of
this article a manifold is simply a surface that is locally Euclidean.  Because
$d^2\omega = 0$, this leads to $ \int_ M d^2\omega = \int_{\partial M} d\omega
= \int_{\partial^2 M} \omega = 0$, where $\partial^2 M$ indicates the boundary
of a boundary of a region. If we assume that $\omega$ is non-vanishing, then
$\partial^2 M$ is $\oslash$. The M$\ddot{o}$bius strip can be used as another
example of the theorem Yang cites.

\noindent
{\bf Gauge Invariance}

\noindent
In a 1918 article, Hermann Weyl\footnote {Weyl, Hermann, (1919), {\it
Sitzwingsber. Preuss. Akad.}, Berlin, 465} tried to combine electromagnetism
and gravity by requiring the theory to be invariant under a local scale change
of the metric $g_{\mu \nu} \rightarrow g_{\mu \nu}e^{\alpha(x)}$, where $x$ is
a 4-vector. This attempt was unsuccessful and was criticized by Einstein for
being inconsistent with observed physical results. It predicted that a vector
parallel transported from point $p$ to $q$ would have a length that was path
dependent. Similarly, the time interval between ticks of a clock would also
depend on the path on which the clock was transported. The article did,
however, introduce\\

\begin{itemize}
  \item The term ``gauge invariance'', his term was {\em Eichinvarianz}. It
    refers to invariance under his scale change. The first use of ``gauge
    invariance'' in English\footnote {See Jackson, J. D. and Okun, L. B.,
    (2001).  {\it Rev. Mod. Physics}, 73, 663.}  was in Weyl's 1929 English
    version\footnote{Weyl, H. (1929). {\it Proc. Natl. Acad. Sci.}, 15, 32.}
    of his famous 1929 paper.

  \item The geometric interpretation of electromagnetism,
  \item The beginnings of non-abelian gauge theory. The similarity of Weyl's
theory to non-abelian gauge theory is more striking in his 1929 paper.

\end{itemize}

\noindent
By 1929 Maxwell's equations had been combined with quantum mechanics to produce
the start of quantum electrodynamics. Weyl in his 1929 article\footnote {Weyl,
Hermann, (1929). {\it Zeit. f. Physic}, 330 56.}  turned from trying to unify
electromagnetism and gravity to following a suggestion originally thought to
have been made by Fritz London in his 1927 article\footnote {London, Fritz,
(1927). {\it Zeit. f. Physic}, 42 375.}, and introduced as a phase factor an
exponential in which the phase $\alpha$ is preceded by the imaginary unit $i$,
e.g., $e^{+iq\alpha({\bf x})}$, in the wave function for the wave equations
(for instance, the Dirac equation is $(i\gamma^\mu \partial_\mu - m)\psi =
0$). It is here that Weyl correctly formulated gauge theory as a symmetry
principle from which electromagnetism could be derived.  It was to become the
driving force in the development of quantum field theory. In their 2001 {\it
Rev. Mod. Phys.} paper Jackson and Okun point out that in a 1926 paper\footnote
{Fock, V., (1926). {\it Z. Phys}, 39 226.}  pre-dating London's, Fock showed
that for a quantum theory of charged particles interacting with the
electromagnetic field, invariance under a gauge transformation of the
potentials required multiplication of the wave function by the now well-know
phase factor.  Many subsequent authors incorrecty cited the date of Fock's
paper as 1927.  Weyl's 1929 article along with his 1918 one and Fock's and
London's, and other key articles appear in translation in a work by
O'Raifeartaigh\footnote {O'Raifeartaigh L., (1997) {\it The Dawning of Gauge
Theory,} Princeton University.} with his comments. Yang\footnote {Yang, C.N.,
(2005) {\it Selected Papers (1945-1980) With Commentary}, p19, World
Scientific.}  discusses Weyl's gauge theory results as reported by
Pauli\footnote{Pauli, W., (1941). {\it Rev. Mod. Physics}, 13, 203.}, as a
source for Yang-Mills gauge theory (although Yang didn't find out until much
later that these were Weyl's results):

\begin{quotation}
\noindent
I was very much impressed with the idea that charge conservation was
related to the invariance of the theory under phase changes, an idea, I later
found out, due originally to H. Weyl. I was even more
impressed with the fact the gauge-invariance determined all the electromagnetic
interactions.
\end{quotation}

\noindent
For the wave equations to be gauge invariant, i.e., have the same form after
the gauge transformation as before, the local phase transformation $\psi({\bf
x}) \rightarrow \psi(x)e^{+iq\alpha({\bf x})}$ has to be accompanied by the
local gauge transformation\\

\noindent
(9) ${\bf A_\mu} \rightarrow {\bf A_\mu} - {\bf \partial_{\mu}\alpha({\bf x})}$\\

\noindent
(The phase and gauge transformations are local because $\alpha({\bf x})$ is a
function of {\bf x}.)  This dictates that the $\partial_\mu$ in the wave
equations be replaced by $\partial_\mu + iqA_\mu$ in order for the ${\bf
\partial_{\mu}\alpha({\bf x})}$ terms to cancel each other. Thus gauge
invariance determines the type of interaction -- here, the inclusion of the
vector potential. This is called the {\it gauge principle} and $A_\mu$ is
called the {\it gauge field} or {\it gauge potential}. Gauge invariance is also
called {\it gauge symmetry.} In electromagnetism, {\bf A} is the space-time
vector potential representing the photon field, while in electroweak theory,
{\bf A} represents the intermediate vector bosons $W^{\pm}$ and $Z^0$ fields
and in the strong interaction, {\bf A} represents the colored gluon fields. The
fact that the $q$ in $\psi(x)e^{+iq\alpha({\bf x})}$ must be the same as the q
in $\partial_\mu + iqA_\mu$ to insure gauge invariance, means that the charge q
must be conserved.\footnote {See section 4.6 of Aitchison, I.J.R., and Hey,
A.J.G., (1989) {\it Gauge Theories in Particle Physics}, Adam Hilger.}  Thus
gauge invariance dictates charge conservation. By Noether's theorem, a
conserved current is associated with a symmetry. Here the symmetry is the
non-physical rotation invariance in an internal space called a {\em fiber}.  In
electromagnetism the rotations form the group U(1), the group of unitary
1-dimensional matrices. U(1) is an example of a {\it structure group} and the
fiber is $S^1$, the circle.  

\noindent
A {\em fiber bundle} is determined by two manifolds and the structure group $G$
which acts on the fiber: the first manifold, called the total space {\it E}
consists of many copies of the fiber {\it F}, one for each point in the second
manifold, the base manifold {\it M} which for our discussion is the space-time
manifold. The fibers are said to project down to the base manifold. A {\it
principal fiber bundle}\footnote{In giving these definitions, we restrict
attention to the smooth manifolds which is adequate for our discussion.} is a
fiber bundle in which the structure group, {\it G}, is a Lie group that acts on
the total space {\it E} in such a way that each fiber is mapped onto itself and
the action of an individual fiber looks like the action of the structure group
on itself by left-translation. In particular, the fiber {\it F} is
diffeomorphic to the structure group $G$.

\noindent
The gauge principle shows how electromagnetism can be introduced into quantum
mechanics.  The transformation $\partial_\mu \rightarrow \partial_\mu + iqA_\mu
$ is also called the {\it minimal principle} and the operation $\partial_\mu +
iqA_\mu$ is the covariant derivative of differential geometry, ${\bf D} = {\bf
d + iqA}$, where {\bf A} is the connection on a fiber bundle.  A connection on
a fiber bundle allows one to identify fibers over points $b_i$ $\epsilon M$ via
parallel transport along a path $\gamma$ from $b_1$ to $b_2$.  In general, the
particular identification is path dependent. It turns out that the parallel
transport depends only on the homotopy class of the path if and only if the
curvature of the connection vanishes identically. Recall that two paths are
homotopic if one can be deformed continuously onto the other keeping the end
points fixed.

\noindent
Weyl in his 1929 paper also includes an expression for the curvature ${\bf
\Omega}$ of the connection {\bf A}, namely Cartan's second structural equation
which in modern differential geometry notation is ${\bf \Omega = dA + A \wedge
A}$.  It is the same form as the equation used by Yang and Mills which in
modern notation is ${\bf \Omega = dA + [A, A]}$, where [] is the Lie bracket.
Since the transformations in (9) form an abelian group U(1), the space-time
vector potential {\bf A} commutes with itself. Thus in electromagnetism the
curvature of the connection {\bf A} is just\\

\noindent
(10) ${\bf \Omega = dA}$.\\

\noindent
which, as we will see in the next section, is the field strength {\bf F}
defined as {\bf F = dA}.

\noindent
{\bf Differential Geometry}

\noindent
Differential geometry principly developed by Levi-Civita, Cartan, Poincar\'{e},
de Rham, Whitney, Hodge, Chern, Steenrod and Ehresmann led to the
develpment of fiber-bundle theory which is used in explaining the geometric
content of Maxwell's equations. It was later used to explain Yang-Mills theory
and to develop string theory. The successes of gauge theory in physics sparked
mathematicians interest in it. In the 1970's Sir Michael Atiyah initiated the
study of the mathematics of the Yang-Mills equations and in 1983 his student
Simon Donaldson using Yang-Mills theory discovered a unique property of smooth
manifolds\footnote{Donaldson, S. K. (1983), {\it Bull. Amer. Math. Soc.} 8,
81.} in $\mathbb{R}^4$.  Michael Freedman went on to prove that there exists
multiple exotic differential structures only on $\mathbb{R}^4$. It is known
that in other dimensions, the standard differential structure on  
$\mathbb{R}^n$ is unique.

\noindent
In 1959 Aharonov and Bohm\footnote{Y. Aharonov and D. Bohm, (1959) {\it
Phys. Rev.} 115, 485.} established the primacy of the vector potential by
proposing an electron diffraction experiment to demonstrate a quantum
mechanical effect: A long solenoid lies behind a wall with two slits and is
positioned between the slits and paralled to them. An electron source in front
of the wall emits electrons that follow two paths. One path through the upper
slit and the other path through the lower slit. The first electron path flows
above the solenoid and the other path flows below it.  The solenoid is small
enough so that when no current flows through it, the solenoid doesn't interfere
with the electrons' flow. The two paths converge and form a diffraction pattern
on a screen behind the solenoid.  When the current is turned on, there is no
magnetic or electric field outside the solenoid so the electrons cannot be
effected by these fields; however there is a vector potential $\vec {\bf A}$
and it effects the interference pattern on the screen. Thus Einstein's
objection to Weyl's 1918 paper can be understood as saying that there is no
Aharonov$-$Bohm effect for gravity. Because of the necessary presence of the
solenoid, the upper path cannot be continuously deformed into the lower
one. Therefore, the two-paths are not homotopically equivalent.

\noindent
The
solution of $(1\slash 2m)(-i\hbar\nabla - q{\bf A}\slash c)^2\psi + qV\psi =
E\psi $, the time-independent Schr$\ddot{o}$dinger's equation for a charged
particle, is $\psi_0({\bf x}) e^{(iq\slash c\hbar)\int^{\bf s(x)} {\bf A(y)}
d{\bf s^{'}(y)} }$ where $\psi_0({\bf x})$ is the solution of the equation for
{\bf A} equals zero and {\bf s(x)} represents each of the two paths. Here $c$
is the speed of light and $\hbar$ is Plank's constant divided by $2\pi$.  The
interference term in the superposition of the solution for the upper path and
that for the lower path produces a difference in the phase of the electron's
wave function called a {\em phase shift}. Here the phase shift is $(q\slash
c\hbar)\oint {\bf A(x)} d{\bf x}$. By Stoke's theorem, the phase shift is
$(q\slash c\hbar)\phi$ where $\phi$ is the magnetic flux in the solenoid, $\int
\vec {\bf B} \cdot d\vec {\bf S}$. Mathematically, their proposal corresponds
to the fact that even if the curvature [the electromagnetic field strength] of
the connection vanishes [as it does outside of the solenoid] parallel transport
along non-homotopic paths can still be path-dependent [producing a shift in the
diffraction pattern].

\noindent
Chambers\footnote{Chambers, R. G., (1960) {\it Phys. Rev. Lett.} 5, 3}
performed an experiment to test the Aharonov and Bohm (A\-B) effect. The
experiment, however, was criticized because of leakage from a tapered magnetic
needle. Tonomura\footnote{Tonomura, Akira, et. al., (1982) {\it
Phys. Rev. Lett.}  48, 1443, and (1986) {\it Phys. Rev. Lett.} 56, 792} et. al.
performed beautiful experiments that indeed verified the A\-B prediction.  Wu
and Yang\footnote{Wu, T. T. and Yang, C. N., (1975) {\it Phys. Rev. D}, 12,
3845} analyzed the prediction of Aharonov and Bohm and comment that different
phase shifts $(q\slash c\hbar)\phi$ may describe the same interference pattern,
whereas the phase factor $e^{(iq\slash c\hbar)\phi}$ provides a unique
description. The equation $e^{2\pi Ni} = 1$, where $N$ is an integer means that
$e^{(iq\slash c\hbar)(\phi + 2\pi Nc\hbar\slash q)}$ = $e^{(iq\slash
c\hbar)\phi}e^{2\pi Ni}$ = $e^{(iq\slash c\hbar)\phi}$. Thus flux of $\phi,
\phi + 2\pi c\hbar\slash q, \phi + 4\pi c\hbar\slash q ...$ all describe the
same interference pattern. Moreover, they introduced a dictionary relating
gauge theory terminology to bundle terminology. For instance, the gauge theory
phase factor corresponds to the bundle parallel transport; and as we shall see,
the Yang-Mills gauge potential corresponds to a connection on a principal fiber
bundle.

\noindent
Let's see how using the primacy of the four-vector potential {\bf A}, we
can derive the homogeneous Maxwell's equations from differential geometry
simply by using the gauge transformation. Then we'll get the non-homogeneous
Maxwell's equations for source-free ({\bf J} = 0) electromagnetism using the
fact that our world is a four-dimensional (space-time) world.

\noindent
We will also show that Maxwell's equations are invariant under the
transformations ${\bf A_\mu} \rightarrow {\bf A_\mu} + {\bf \partial_{\mu}
\alpha({\bf x})}$, or expressed in differential geometry terms, ${\bf A}
\rightarrow {\bf A} + {\bf d \alpha({\bf x})}$.  We want $\alpha({\bf x})$ to
vanish when a function of {\bf A} is assigned to the $\vec {\bf E}$ and $\vec
{\bf B}$ fields. Taking the exterior derivative of {\bf A} will do this since
${\bf d^2\alpha({\bf x}) = 0}$. Set {\bf A} to the 1-form {\bf A} = -$A_0{\bf
d}t + A_x{\bf d}x + A_y{\bf d}y + A_z{\bf d}z$. Evaluating {\bf dA} and
realizing that the wedge product ${\bf d}x^i \wedge {\bf d}x^j = - {\bf d}x^j
\wedge {\bf d}x^i$ and therefore ${\bf d}x^j \wedge {\bf d}x^j = 0$ where ${\bf
d}x^0$ is ${\bf d}t$, ${\bf d}x^1$ is ${\bf d}x$, ${\bf d}x^2$ is ${\bf d}y$
and ${\bf d}x^3$ is ${\bf d}z$, produces the 2-form {\bf dA} consisting of
terms like $(\partial_xA_0 + \partial_tA_x)dtdx$ and $(\partial_x A_y -
\partial_y A_x)dxdy$.  When all the components are evaluated, these terms
become respectively $\nabla {\bf A_0} + \partial \vec {\bf A}/ \partial_t$ and
$\nabla \times \vec {\bf A}$. The analysis up to now has been purely
mathematical. To give it physical significance we associate these terms with
the field strengths $\vec {\bf B}$ and $\vec {\bf E}$. In electromagnetic
theory, two fundamental principles are $\nabla \cdot \vec {\bf B}= 0$ (no
magnetic monopoles) and for time-independent fields $\vec {\bf E} = - \nabla
A_0$ (the electromagnetic field is the gradient of the scalar potential), so
consistency dictates that in the time-dependent case, we assign the two terms
to $\vec {\bf B}$ and $\vec {\bf E}$ respectively:\\

\noindent
(11) $\vec {\bf B} = \nabla \times \vec {\bf A}$ and 
$\vec {\bf E} = - \nabla A_0 - \partial_t \vec {\bf A}$\\

\noindent
The gradient, curl and divergence are spatial operators -- they envolve the
differentials dx, dy and dz. The exterior derivative of a scalar is the
gradient, the exterior derivative of a spatial 1-form is the curl, and the
exterior derivative of a spatial two-form is the divergence. In the 1-form {\bf
A}, the -$A_0{\bf d}t$ is a spatial scalar and when the exterior derivative is
applied gives rise to $\nabla A_0$. The remaining terms in {\bf A} are the
coefficients of $dx^i$ constituting a spatial 1-form and thus produce $\nabla
\times \vec {\bf A}$.

\noindent
We define the field strength, {\bf F} as {\bf F = dA} and from equation (10),
we see that the field strength is the curvature of the connection {\bf
A}. Using the equations in (11) and the 2-form {\bf dA} we get\\

\noindent
(12) {\bf F} = $E_x{\bf d}x{\bf d}t + E_y{\bf d}y{\bf d}t + E_z{\bf d}z{\bf d}t
+ B_x{\bf d}y{\bf d}z + B_y{\bf d}z{\bf d}x + B_z{\bf d}x{\bf d}y$\\

\noindent
where for example ${\bf d}x{\bf d}t$ is the wedge product ${\bf d}x \wedge {\bf
d}t$. Since ${\bf d^2A = 0}$\\

\noindent
(13) ${\bf dF = 0}$\\

\noindent
Evaluating {\bf dF} gives the homogeneous Maxwell's equations. In equation (12)
since the {\bf E} part is a spatial 1-form, when the exterior derivative is
applied, it produces the $\nabla \times \vec {\bf E}$ part of Maxwell's
homogenous equations. Since the {\bf B} part of equation (12) is a spatial
2-form, it results in the $\nabla \cdot \vec {\bf B}$ part.  Since ${\bf dF =
0}$, {\bf F} is said to be a closed 2-form.

\noindent
To get the expression for the non-homogeneous Maxwell's equations, i.e., the
equivalent of equation (7),  we use\\

\noindent
(14) {\bf J} = $\rho{\bf d}t + J_x{\bf d}x + J_y{\bf d}y + J_z{\bf d}z$\\

\noindent
and calculate the Hodge dual using the Hodge star operator.  The Hodge Duals
are defined\footnote {Misner, C. W., Thorne, K. S., and Wheeler, J. A., (1973)
{\it Gravitation}. Freeman, San Francisco.}  by $^*F_{\alpha \beta} =
1/2\epsilon _{\alpha \beta \gamma \delta} F^{\gamma \delta}$ and $^*J_{\alpha
\beta \gamma} = \epsilon _{\alpha \beta \gamma \delta} J^{\delta}$.  The Hodge
star\footnote {Flanders, H., (1963).  {\it Differential Forms}. Academic
Press.} operates on the differentials in equation (12) and (14) using
$^*(dx^idt) = dx^jdx^k$ and $^*(dx^jdx^k) = - dx^idt$ where i, j, and k refer
to x, y and z, and are taken in cyclic order. The metric used is (-+++). Thus
the Hodge star takes a spatial 1-form $dx^idt$ into a spatial 2-form and vice
versa with a sign change.

\noindent
The non-homogeneous Maxwell's equations are then expressed by\\

\noindent
(15) {\bf d*F} = 0 (source-free)\\

\noindent
(15') {\bf d*F} = {\bf *J} (non-source-free)\\

\noindent
where the 2-form {\bf *F} and the 3-form {\bf *J} are respectively the Hodge
duals of {\bf F} and {\bf J}. {\bf *F} and {\bf *J} are defined as\\

\noindent
(16) {\bf *F} = $-B_x{\bf d}x{\bf d}t -B_y{\bf d}y{\bf d}t -B_z{\bf d}z{\bf d}t
+ E_x{\bf d}y{\bf d}z + E_y{\bf d}z{\bf d}x + E_z{\bf d}x{\bf d}y$\\

\noindent
(17) {\bf *J} = $\rho{\bf d}x{\bf d}y{\bf d}z - J_x{\bf d}t{\bf d}y{\bf d}z -
J_y{\bf d}t{\bf d}z{\bf d}x - J_z{\bf d}t{\bf d}x{\bf d}y$\\

\noindent
Thus the Hodge star reverses the rolls of $\vec {\bf E}$ and $\vec {\bf B}$
from what they were in {\bf F}. In {\bf *F} the coefficient of the spatial
1-form is now $-\vec {\bf B}$ which will produce the curl in the
non-homogeneous Maxwell's equations, and the coefficient of the spatial 2-form
is $\vec {\bf E}$ which will produce the divergence.  In $\mathbb{R}^n$, the
Hodge star operation on a p-form produces an (n-p)-form. Thus the form of
Maxwell's equations is dictated by the fact that we live in a four-dimensional
world.  When the 1-form {\bf A} undergoes the local gauge transformation ${\bf
A} \rightarrow {\bf A} + {\bf d\alpha({\bf x})}$, {\bf dA} remains the same
since ${\bf d^2\alpha} = 0$. Since $\vec {\bf B}$ and $\vec {\bf E}$ are
unchanged, Maxwell's theory is gauge invariant.

\noindent
{\bf The Dirac and Electromagnetism Lagrangians}\\

\noindent
To prepare for the discussion of the Yang-Mills equations, let's investigate
the Dirac and Electromagnetism Lagrangians The Dirac equation is \\

(20) $(i\gamma^\mu \partial_\mu - m)\psi = 0$\\

\noindent
where the speed of light, $c$, and Plank's constant $\hbar$ are set to one.
Its Lagrangian density is\\

(21) ${\mathcal L} = \bar{\psi}(i\gamma^\mu \partial_\mu - m)\psi$\\

\noindent
The Euler-Lagrange equations minimize the action {\bf S} where $S = \int
{\mathcal L} {\bf dx}$. Using the Euler-Lagrange equation where the
differentiation is with respect to $\bar{\psi}$, i.e.,\\

(22) $ \partial_{\mu} {\textstyle
( \frac {\partial {\mathcal L}} {\partial(\partial_\mu
  \bar {\psi})})- \frac {\partial {\mathcal L}} {\partial \bar{\psi}}} = 0$\\

\noindent
yields equation (20).

\noindent
The same gauge invariant argument used in the Gauge Invariance section applies
here. In order for the Lagrangian to be invariant under the phase
transformation $\psi({\bf x}) \rightarrow \psi(x)e^{+i\alpha({\bf x})}$, this
transformation has to be accompanied by the local gauge transformation ${\bf
A_\mu} \rightarrow {\bf A_\mu} - e^{-1}{\bf \partial_{\mu}\alpha({\bf x})}$ and
$\partial_\mu$ has to be replaced by $\partial_\mu + ieA_\mu$. The Lagrangian
density becomes\\

\noindent
(23)  ${\mathcal L} = \bar{\psi}(i\gamma^\mu \partial_\mu - m)\psi -
e \bar{\psi}\gamma^\mu \psi A_\mu $\\

\noindent
The last term is the equivalent of the interaction energy with the
electromagnetic field, $j^\mu A_\mu$. In order for the Euler-Lagrangian
equation differentiated with respect to $A_\mu$ to yield the inhomogeneous
Maxwell equation (7) we must add $-(\frac {1} {4})(F_{\mu \nu})^2$ getting\\

(24) ${\mathcal L} = \bar{\psi}(i\gamma^\mu \partial_\mu - m)\psi -
e \bar{\psi}\gamma^\mu \psi A_\mu -(\frac {1} {4})(F_{\mu \nu})^2$ \\

\noindent
The Euler-Lagrange equation yields\\

(25)  $\partial_{\mu} F^{\mu \nu} = e \bar{\psi}\gamma^\nu \psi$\\

\noindent
which equals $J^{\nu}$.  Note that the gauge field $A_\mu$ does not carry a
charge and there is no gauge field self-coupling which would be indicated by an
$[A_\mu, A_\nu]$ term in (25). The Lagrangian density does not yield the
homogeneous Maxwell equations. They are satisfied trivially because the
definition of $F^{\mu \nu}$ satisfies the homogeneous equations
automatically.\footnote {Jackson, J. D., (1998).  {\it Classical
Electrodynamics, 3rd Ed, p600}. John Wiley and Sons.}\\

\noindent
From this it is apparent that the Lagrangian density for the electromagnetic
field alone\\

(26) ${\mathcal L} =  -J^\mu A_\mu -(\frac {1} {4})(F_{\mu \nu})^2$ \\

\noindent
yields all of Maxwell's equations.\\

\noindent
In differential geometry, if $j = 0$, this Lagrangian density becomes\\

(27) ${\mathcal L} =  (\frac {1} {2}F \wedge *F)$ \\

\noindent
{\bf The Yang-Mills Theory}

\noindent
The Yang-Mills theory incorporates isotopic spin symmetry introduced in 1932 by
Heisenberg who observed that the proton and neutron masses are almost the same
(938.272 MeV versus 939.566 MeV respectively). He hypothesized that if the
electromagnetic field was turned off, the masses would be equal and the proton
and neutron would react identically to the strong force, the force that binds
the nucleus together and is responsible for the formation of new particles and
the rapid (typically their lifetimes are about$10^{-20}$ seconds) decay of
others. In a non-physical space (also known as an internal space) called {\it
isospin space}, the proton would have isospin up, for instance, and the
neutron, isospin down; but other than that, they would be identical. The wave
function for each particle could be transformed to that for the other by a
rotation using the spin matrices of the non-abelian group SU(2).  Because of
charge independence, the strong interactions are invariant under rotations in
isospin space.  Since the ratio of the electromagnetic to strong force is
approximately $\alpha$, where $\alpha = e^2/4\pi\hbar c$ = 1/137, to a good
approximation we can neglect the fact that the electromagnetic forces break
this symmetry.  By Noether's theorem, if there is a rotational symmetry in
isospin space, the total isotopic spin is conserved. This hypothesis enables us
to estimate relative rates of the strong interactions in which the final state
has a given isospin. The spin matrices turn out to be the Pauli matrices
$\sigma_i$. The theory just described is a global one, i.e., the isotopic spin
is independent of the space-time coordinate and thus no connection is used. We
will see that Yang and Mills\footnote {Yang, C. N. and Mills, R. L., (1954).
{\it Phys.  Rev.}  96, p91} elevated this global theory to a local one.  In
1954 they proposed applying the isospin matrices to electromagnetic theory in
order to describe the strong interactions. Ultimately their theory was used to
describe the interaction of quarks in the electroweak theory\footnote{The weak
and electromagnetic forces are the two manifestations of the electroweak force}
and the gluons fields of the strong force. In the next section
we will give an example using
the up quark $u$ which has a charge of $\frac {2} {3}e$ and down quark $d$
which has a charge of -$\frac {1} {3}e$.

\noindent
We have seen that the field strength (which is also the curvature of the
connection on the fiber) is given by ${\bf F = dA + A \wedge A}$. In
electromagnetism {\bf A} is a 1-form with scalar coefficients for ${\bf d}x^i$
so ${\bf A \wedge A}$ vanishes. If, however, the coefficients are non-commuting
matrices ${\bf A \wedge A}$ does not vanish and provides for gauge field
self-coupling.  Yang and Mills formulated the field strength, using the letter
$B$ instead if $A$, so we will follow suit. The field is\\

\noindent
(28) $F_{\mu \nu} = (\partial_{\nu}B_{\mu} - \partial_{\mu}B_{\nu}) +i\epsilon
(B_\mu B_\nu - B_\nu B_\mu)$ \\

\noindent
or equivalently $F_{\mu \nu} = (\partial_{\mu}B_{\nu} - \partial_{\nu}B_{\mu})
+i\epsilon [B_\mu, B_\nu]$, where $B$ is the connection on a principal fiber
bundle, i.e., the gauge potential. So as opposed to the electromagnetic field
strength which is linear, their field strength is non-linear.  They proposed
using a local phase. For instance, one could let\\

\noindent
(29) $\psi({\bf x}) \rightarrow \psi(x)e^{i\alpha_j({\bf x})\sigma^j}$\\

\noindent
where $\sigma^j$ are the Pauli matrices and $j$ goes from 1 to 3. Thus the
 exponent includes the dot product (or inner product) in $\mathbb{R}^3$. The
 Pauli matrices do not commute, $[\frac {\sigma^i}{2}, \frac {\sigma^j}{2}] = i
 \epsilon^{ijk} \frac{\sigma^k}{2}.$ Since $B_\mu = \frac {1}{2}b^i_\mu
 \sigma_i$ or $B_\mu = \frac {1}{2}\vec {\bf b}\cdot \vec{\bf \sigma}$ (where
 $b^i_\mu$ is called the isotopic spin vector gauge field) the four-vectors
 $B_\mu$ and $B_\nu$ in (28) do not commute.  The purpose of the Pauli spin
 matrices in the connection $B$ is to rotate the particles in isospin space so
 that they retain their identities at different points in $\mathbb{R}^4$.
 Equation (28) can be rewritten so that the curvature is defined as ${\bf F =
 dB + i\epsilon [B, B]}$. As opposed to the Maxwell's equations case, the
 exterior derivative of the curvature ${\bf dF}$, does not equal zero because
 of the commutator in the expression for the curvature. Thus the exterior
 derivative for the 2-form ${\bf F}$ has to be altered to include the
 connection {\bf B}.

\noindent
The Lagrangian\\ 

\noindent
(30) ${\mathcal L} = \bar{\psi}(i\gamma^\mu (D_\mu - m)\psi - (\frac {1}
     {4})Tr(F_{\mu \nu}F^{\mu \nu})$\\

\noindent
is invariant under the gauge transformation for the covariant derivative
given as\\

\noindent
(32) $D_\mu = \partial_{\mu} -i\epsilon B_\mu$

\noindent
where $\epsilon$ is the coupling constant analagous to $q$ in (9).
The connection $B_\mu$ transforms as\\

\noindent
(33) ${B_\mu} \rightarrow {B_\mu} +\epsilon^{-1} \partial_{\mu}{\bf
  \alpha} + [{\bf \alpha}, {B_\mu}]$,\\

\noindent
the fiber is the sphere, $S^2$ and the structure group is SU(2). 

\noindent
Since there are three components of the vector gauge field $b^i_\mu$, there are
three vector gauge fields representing three gauge particles having spin one.
They were later identified as the intermediate vector bosons $W^{\pm}$ and
$Z^0$ which mediate the electroweak interactions. The fact that there are three
gauge particles is dictated by the fact that the gauge field is coupled with
the three Pauli spin matrices.  Also, since the charges of the up quark and
down quark differ by one, the gauge field particles that are absorbed and
emitted by them in quark-quark interactions can have charges of $\pm$1 or
zero. It's astonishing that Yang and Mills in their 1954 paper predicted the
existence of the three intermediate vector bosons.

\noindent
The gauge particles predicted by the Lagrangian (30) have zero mass since any
mass term added to (30) would make the Lagrangian non-invariant under a local
gauge transformation. So the force associated with the particles would have
infinite range as the photons of the electromagnetic interaction do. Of course
the weak force (the force responsible for particle decaying slowly, typically
their lifetimes are about $10^{-10}$ seconds or much less) and strong nuclear
force are short range. This discrepency was corrected some years later by the
introduction of spontaneous symmetry breaking in the electroweak $SU(2) \times
U(1)$ theory of Weinberg, Salam and Glashow (WSG) using the Higgs
mechanism. The WSG theory, which explains the electromagnetic and weak forces,
predicts the existence of four gauge bosons: the three massive ones, $W^\pm$
and $Z^0$, and the photon. Moreover, it predicts the mass of the $W^\pm$ (80.37
$\pm$0.03 GeV) and $Z^0$ (92 $\pm$ 2 GeV), where GeV represents a billion
electron volts.  The $W^\pm$ was discovered\footnote {Arnison, G. et. al.,
(1983).  {\it Phys. Lett.} 122B, 103} in 1983 (its mass is now reported at
80.425 Gev $\pm$ 0.033 GeV) and later that year the $Z^0$ was
discovered\footnote {Arnison, G. et. al., (1983).  {\it Phys. Lett.} 126B, 398}
(its mass is now reported at a mass of 91.187 $\pm$ 0.002 GeV).

\noindent
The Euler-Lagrange equations for equation (30) give the Dirac equation\\

\noindent
(34) $\gamma^\mu (\partial_\mu -ie B_\mu)\psi + m\psi = 0$\\

\noindent
and also the vector equation for the vector field {\bf F}, namely\\

\noindent
(35) $\partial^{\mu} {\bf F}_{\mu \nu} - i \epsilon [{\bf B}^\mu, {\bf F}_{\mu
     \nu}] = - \frac {1} {2}\epsilon \bar{\psi}\gamma_\nu {\bf \sigma}\psi
= -{\bf J}_\nu$\\

\noindent
which, if it weren't for the commutator, is the same form as the
non-homogeneous four-vector Maxwell equation. The commutator
causes the gauge particles to interact with themselves.

\noindent
The effect of these equation is explained by 't Hooft
\footnote{ 't Hooft, Gerardus, editor, (2005) {\it 50 Years of Yang-Mills
Theory}, World Scientific.} who with Veltman proved the renormalizability of
Yang-Mills theories.

\begin{quotation}
\noindent
...The {\it B} quanta would be expected to be exchanged between any pair of
particles carrying isospin, generating not only a force much like the
electro-magnetic force, but also a force that rotates these particles in
isospin space, which means that elementary reactions envolving the
transmutation of particles into their isospin partners will result. A novelty
in the Yang-Mills theory was that the {\it B} quanta are predicted to interact
directly with one another. These interactions originate from the commutator
term in the ${\bf F}_{\mu \nu}$ field [equation (35)], but one can understand
physically why such interactions have to occur: in contrast with ordinary
photons, the Yang-Mills quanta also carry isospin, so they will undergo isospin
transitions themselves, and furthermore, some of them are charged, so the
neutral components of the Yang-Mills fields cause Coloumb-like interactions
between these charged particles.
\end{quotation}

\noindent
So the Yang-Mills equations indicate that for instance for the up quark down
quark doublet, the $W^-$ generates a force that rotates the $d$ into the $u$ in
isospin space exhibited by the transition $d \rightarrow u + W^-$. The
commutator in equation (35) is responsible for interactions like $W \rightarrow
W + Z$ occurring\footnote{This is indicated in Figure 1 of the Yang-Mills
paper}, and the $W$ can radiate producing a photon in $W \rightarrow W +
\gamma$.

\noindent
The Yang-Mills equations can be derived from the differential geometry
Lagrangian density, where $k$ is a constant\\

\noindent
(36) ${\mathcal L} = -k Tr(F \wedge *F)$.\\

\noindent
The Euler-Lagrange equations produce $d_B F =0$ (the Bianchi identity) and in
the absence of currents, $d_B *F =0$ where $d_B$ is the exterior covariant
derivative. These are the Yang-Mills equations in compact form. 

\noindent
{\bf The Feynman Stamp}\\
\noindent
In QED after Schwinger, Tomonaga and Feynman addressed the singularites
produced by the self-energy of the electron by renormalizing the theory, they
were then exceedingly successful in predicting phenomena such as the Lamb
shift and anomalous magnetic moment of the electron.\\

\noindent  
Feynman introduced\footnote{Feynman, R. P., (1949), {\it Phys. Rev.} 76, 769}
schematic diagrams, today called {\it Feynman diagrams}, to facilitate
calculations of particle interaction parameters. External particles,
represented by lines (edges) connected to only one vertex are real, i.e.,
observable. They are said to be on the mass shell, meaning their four-momentum
squared equals their actual mass, i.e., $m^2 = E^2 -p^2$. Internal particles
are represented by lines that connect vertices and are therefore intermediate
states -- that's why they are said to {\it mediate} the interaction.  They are
virtual and are considered to be off the mass shell.  This means their
four-momentum squared differs from the value of their actual mass. This is done
so that four-momentum is conserved at each vertex.  The rationale for this
difference is the application of the uncertainty principle $\Delta E \cdot
\Delta t = \hbar$. Since $\Delta t$, the time spent between external states is
very small, for that short time period, $\Delta E$ and thus the difference
between the actual and calculated mass can be large. In the following Feynman
diagrams, the time axis is vertical upwards.

\noindent
The diagram on the upper-left of the stamp (Figure 1) is a vertex diagram, and
as such represents a component of a Feynman diagram. It illustrates the
creation of an electron-positron pair from a photon, $\gamma$; it's called {\it
pair production.}  The $\gamma$ is represented by a wavy line.  The
Feynman-Stuckelberg interpretation of negative-energy solutions indicates that
here the positron, the electron's antiparticle, which is propagating forward in
time is in all ways equivalent to an electron going backwards in time. If all
the particles here were external, the process would not conserve energy and
momentum. To see this you must first remember that since the photon has zero
mass -- due to the gauge invariance of electromagnetic theory-- its energy and
momentum are equal. Thus $\beta$ which equals $\frac {p} {E}$ has the value 1;
but $\beta$ = $\frac {v} {c}$ so that the photon's velocity is always c, the
speed of light. In the electron-positron center of mass frame (more aptly
called the center of momentum frame, since the net momentum of all the
particles is zero there), the electron and positron momenta are equal and are
in opposite directions. The photon travels at the speed of light and therefore
its momentum cannot be zero; but there is no particle to cancel its momentum,
so the interaction cannot occur (for it to occur requires a Coloumb field from
a nearby nucleus to provide a virtual photon that transfers momentum producing
a nuclear recoil). Therefore the $\gamma$ in the diagram is internal. Its mass
is off the mass shell and cannot equal its normal value, i.e., zero.

\noindent
The diagram on the lower-left of the stamp (Figure 2) is also a vertex diagram
and represents an electron-positron pair annihilation producing a
$\gamma$. Again, if all the particles are external, conservation energy and
momentum prohibits the reaction from occurring, So the $\gamma$ must be
virtual.

\noindent
The diagram (Figure 3) on the bottom to the right of Feynman was meant to
represent an electron-electron scattering with a single photon exchange. This
is called M{\o}ller scattering. (It can, however, represent any number of
interactions exchanging a photon.) The diagram represents the $t$-channel of
M{\o}ller scattering; there is another diagram not shown here representing the
$u$-channel contribution where $u$, $t$ and another variable $s$ are called the
{\it Mandelstam variables}. They are used in general to describe 2-body
$\rightarrow$ 2-body interactions. If you rotate the diagram in Figure 3 by
$90^o$ you have the $s$-channel diagram for electron-positron scattering called
Bhabha scattering shown in Figure 4 but not on the stamp. Here an electron and
positron annihilate producing a virtual photon which in turn produces an
electron-positron pair. There is also a $t$-channel contribution to Bhabha
scattering. The cross section for Bhabha scattering can be easilly obtained
from the one for M{\o}ller scattering by interchanging the $s$ and $u$ in the
cross section expression in a process called {\it crossing}. Small angle Bhabha
scattering is used to test the luminosity in $e^+$-$e^-$ colliding beam
accelerators.

\noindent
To the right of the M{\o}ller scattering diagram is a vertex correction to
electron scattering shown in Figure 5 where the extra photon forms a loop. It
is used to calculate both the anomalous magnetic moment of the electron and
muon, also the anomalous magnetic moment contribution to the Lamb
shift\footnote{See for instance p156, Griffith, David, (1987) {\it Introduction
to Elementary Particles} John Wiley and Sons.}. The other two contributions to
the Lamb shift are the vacuum polarization and the electron mass
renormalization.  The Lamb shift explains the splitting in the spectrum of the
$2S_ \frac {1} {2}$ and $2P_ \frac {1} {2}$ levels of hydrogen; whereas Dirac
theory alone incorrectly predicted that these two levels should be degenerate.

\noindent
The low-order solution of the Dirac equation predicts a value of 2 for the
g-factor used in the expression for the magnetic moment of the electron. The
vertex correction shown in Figure 5, however, alters the g-factor producing an
anomalous magnetic moment contribution written as $\frac {g - 2} {2}$. When
this and higher order contributions are included, the calculated value of
$\frac {g - 2} {2}$ for the electron is 1159 652 460(127)(75) $\times$
$10^{-12}$ and the experimental value is 1159 652 193(10) $\times$ $10^{-12}$
where the numbers in parenthesis are the errors. This seven-significant figure
agreement is a spectacular triumph for QED.  We need not emphasise that the
calculations for all these diagrams use the gauge principal for quantum
electrodynamics.

\noindent 
The other diagrams on the stamp are all vertex diagrams and show how Feynman's
work originally applicable to QED was then later used to elucidate the
electroweak force. This is exemplified on the stamp by flavor changing
transitions, e.g., $d \rightarrow W^- + u $ shown in Figure 6 and flavor
conserving transitions, e.g., $d \rightarrow Z^0 + d $ of the electroweak force
-- the $u$ and $d$ quarks have different values of flavor. The process in
Figure 6 occurs for instance in $\beta$ decay where a neutron (udd) decays into
a proton (udu) and electron and an anti-neutrino. What happens is that the
transition $d \rightarrow u + W^-$ corresponds to a rotation in isospin
space. This rotation is caused by the virtual $W^-$ which mediates the
decay. It in turn decays into an electron and an anti-neutrino. The
calculations for these transitions all use the Yang-Mills equations. Although
the quarks are confined in the hadrons -- particles that undergo strong
interactions like the proton and neutron -- they are free to interact with the
intermediate vector bosons.

\noindent
{\bf Who Designed The Stamp?}\\

\noindent
Feynman's daughter Michelle was sent a provisional version of the stamp by the
United States Postal Service and advised on the design of the stamp by among
others, Ralph Leighton, coauthor with Richard Feynman of two popular books; and
Cal Tech's Steven Frautschi and Kip Thorne. Frautschi and Leighton edited the
Feynman diagrams, and Frautschi rearranged them and composed the final design.
The person who chose the original Feynman diagrams that form the basis for the
stamp remains a mystery.

\noindent
{\bf Acknowledgements}\\

\noindent 
The author thanks Jeff Cheeger, Bob Ehrlich, J. D. Jackson and C. N. Yang, for
their advice on the article.

\noindent
{\bf Appendix A: Yang-Mills Derivation}

\noindent
We begin by performing a phase transformation

\noindent
(A1) $\psi ' = S\psi$\\ 

\noindent
where $S = e^{i{\bf \alpha(x) \cdot \sigma}}$ and use the covariant derivative
$D_\mu = \partial_\mu -i\epsilon B_\mu$ which transforms in the same way as
indicated in equation (A2)\\

\noindent
(A2) $D'\psi' = SD\psi$. Then\\

\noindent
(A3) $(\partial_\mu -i\epsilon B'_\mu)S\psi= (\partial_\mu S)\psi + 
S\partial_\mu \psi -i\epsilon B'_\mu \psi$\\

\noindent
But (A3) equals $S\partial_\mu\psi -i\epsilon SB_\mu \psi.$\\

\noindent
Cancelling $S\partial_\mu\psi$ on both sides we get,\\ 

\noindent
(A4) $(\partial_\mu S) \psi -i\epsilon B'_\mu S\psi = -i\epsilon SB_\mu \psi$,
or\\

\noindent
(A5) $-i\epsilon B'_\mu S = -i\epsilon SB_\mu - (\partial_\mu S) $
or $ B'_\mu S = SB_\mu + (\partial_\mu S)/(i\epsilon) $, thus \\

\noindent
(A6) $ B'_\mu S = SB_\mu -i(\partial_\mu S)/\epsilon $ or\\

\noindent
(A7) $ B'_\mu = SB_\mu S^{-1} -i(\partial_\mu S)S^{-1}/\epsilon $\\

\noindent
For $\alpha$ infinitesimal,  $S = 1 + i\alpha \cdot \sigma$, so\\

\noindent
(A8) $B'_\mu = (1 + i\alpha \cdot \sigma)B_\mu (1 - i\alpha \cdot \sigma ) -
i(1/\epsilon) \partial_\mu (1 + i\alpha \cdot \sigma ) (1 - i\alpha \cdot
\sigma)$ and \\

\noindent
Remembering that $(a \cdot \sigma)(b \cdot \sigma) = a \cdot b + i\sigma \cdot
(a \times b)$, setting $B_\mu = \sigma \cdot b_\mu$, and since $\alpha$ is
infintessimal, we drop terms of order $\alpha^2$ getting\\

\noindent
(A9) $b'_\mu \cdot \sigma = b_\mu \cdot \sigma + i[(\alpha \cdot \sigma)(b_\mu
  \cdot \sigma), (b_\mu \cdot \sigma)(\alpha \cdot \sigma)] + (1/\epsilon)
\partial_\mu (\alpha \cdot \sigma )$ and finally\\

\noindent
(A10) $b'_\mu = b_\mu + 2(b_\mu \times \alpha) + (1/\epsilon) \partial_\mu
\alpha$, which is equation (10) in the Yang-Mills paper. \\

\noindent
Pauli, in equation (22a) of Part I of his 1941 Rev. Mod. Phys. article gives
the electromagnetic field strength as $[D_\mu, D_\nu] = -i\epsilon F_{\mu \nu}$
which apart from the minus sign agrees with our conventions and where $D_\mu =
\partial_\mu -i\epsilon A_\mu$.  So by following suit, the field strength for
the Yang-Mills strength can be obtained from the commutator\\

\noindent
(A11) $[D_\mu, D_\nu] = (\partial_\mu - i\epsilon B_\mu)(\partial_\nu -
i\epsilon B_\nu) - (\partial_\nu - i\epsilon B_\nu)(\partial_\mu - i\epsilon
B_\mu)$\\

\noindent
operating on the wave function $\psi$.  Note that $ -\partial_\mu (B_\nu \psi)
= -(\partial_\mu B_\nu) \psi - B_\nu \partial_\mu \psi$. So we get an
apparently extra $- B_\nu \partial_\mu$ and a $B_\mu \partial_\nu$ term. Thus
expanding (A11) we get\\

\noindent
(A12) $\partial_\mu \partial_\nu - i\epsilon \partial_\mu B_\nu - i\epsilon
B_\mu \partial_\nu - i\epsilon B_\nu \partial_\mu - \epsilon^2 B_\mu B_\nu -
\partial_\nu \partial_\mu + i\epsilon \partial_\nu B_\mu + i\epsilon B_\nu
\partial_\mu + i\epsilon B_\mu \partial_\nu + \epsilon^2 B_\nu B_\mu $\\

\noindent
which reduces to\\

\noindent
(A13) $i\epsilon (\partial_\nu B_\mu - \partial_\mu B_\nu) - \epsilon^2
[B_\mu, B_\nu]$ or\\

\noindent
(A14) $[D_\mu, D_\nu] =  i\epsilon F_{\mu \nu}$ where $F_{\mu \nu}$ is given by
equation (28).\\

\noindent
If we let $B_\mu = \sigma \cdot b_\mu$ we can rewrite the equation 
$F_{\mu \nu} = (\partial_{\nu}B_{\mu} - \partial_{\mu}B_{\nu}) +i\epsilon
(B_\mu B_\nu - B_\nu B_\mu)$ as\\

\noindent
(A15)
$F_{\mu \nu} = (\partial_{\nu}B_{\mu} - \partial_{\mu}B_{\nu}) + i\epsilon
(2i\sigma \cdot b_{\mu} \times b_{\nu})$\\

\noindent
If we further let $F_{\mu \nu} = f_{\mu \nu} \cdot \sigma$, we get\\

\noindent
(A16)
$f_{\mu \nu} = (\partial_{\nu}b_{\mu} - \partial_{\mu}b_{\nu}) - 2 \epsilon
b_{\mu} \times b_{\nu}$\\

\noindent
which is equation (9) in the Yang-Mills paper.

\noindent
{\bf Appendix B, Finding the Field Strength}

\noindent
We reconstruct how one can go about determining the field strength. Since

\begin{equation}F'_{\mu \nu} = S^{-1}F_{\mu \nu}S\end{equation}

\noindent
under an isotopic gaude transformation, let's start off with the
electromagnetic-like field strength in the primed syst\ em

\begin{equation}F'_{\mu \nu} = \partial_{\nu}B'_{\mu} -\partial_{\mu}B'_{\nu}
\label{eq:F} \end{equation}

\noindent
and express it in terms of the non-primed system fields. We
calculate $\partial_{\nu}B'_{\mu}$ from $B'_\mu = S^{-1}B_\mu S +
iS^{-1}(\partial_\mu S)/\epsilon$, equation (A7), obtaining

\begin{center}$\partial_{\nu}B'_{\mu}=-S^{-1}(\partial_{\nu}S)S^{-1}B_{\mu}S+
S^{-1}(\partial_{\nu}B_{\mu})S+S^{-1}B_{\mu}\partial_{\nu}S$
+\end{center}
\begin{equation}
i/\epsilon[-S^{-1}(\partial_{\nu}S)S^{-1}\partial_{\mu}S+
S^{-1}\partial_{\nu}\partial_{\mu}S]
\end{equation}

\noindent
So

\begin{center}
$\partial_{\nu}B'_{\mu} -\partial_{\mu}B'_{\nu} = -S^{-1}
[(\partial_{\nu}S)S^{-1}B_{\mu}- (\partial_{\mu}S)S^{-1}B_{\nu}]S$
\end{center}
\begin{center}
$+S^{-1}[\partial_{\nu}B_{\mu}- \partial_{\mu}B_{\nu}]S+
S^{-1}[B_{\mu}\partial_{\nu}- B_{\nu}\partial_{\mu}]S+$
\end{center}

\begin{equation}i/\epsilon[-S^{-1}(\partial_{\nu}S)S^{-1}\partial_{\mu}S +
S^{-1}(\partial_{\mu}S)S^{-1}\partial_{\nu}S]\end{equation}

\noindent
We see that the $+S^{-1}[\partial_{\nu}B_{\mu}- \partial_{\mu}B_{\nu}]S$ term
satisfies equation (1) if the field strength only had the electromagnetic-like
contribution. The other terms must either represent the transformed
non-electromagnetic-like part of $F_{\mu \nu}$ or be cancelled by adding the
non-electromagnetic terms to equation (2). Since $S$ is only used for the
transformation, it should not appear in the expression for $F_{\mu \nu}$.

\noindent
The $i/\epsilon$ term in equations (4) dictates that a term multiplied by
$i\epsilon$ be added to equation (2).  Since $S^{-1}(\partial_{\mu}S)$ and
$S^{-1}\partial_{\nu}S$ appear in the expressions for $B'_\mu$ and $B'_\nu$
respectively, the product of $S^{-1}(\partial_{\mu}S)$ and
$S^{-1}\partial_{\nu}S$ that appears in the last term of equation (4) suggests
that we should start our quest to eliminate extra terms in equation
(4) by adding $i\epsilon B'_\mu B'_\nu$ to that equation. This product gives

\begin{center}
$i\epsilon B'_\mu B'_\nu = i\epsilon[S^{-1}B_\mu S +
iS^{-1}(\partial_\mu S)/\epsilon]*$
\end{center}
\begin{equation}
[ S^{-1}B_\nu S + iS^{-1}(\partial_\nu S)/\epsilon]\end{equation}

\noindent
which equals
\\
\\
\begin{center}
$i\epsilon S^{-1} B_\mu B_\nu S -i/\epsilon S^{-1}(\partial_\mu S)S^{-1}
\partial_\nu S -$
\end{center}
\begin{equation}
S^{-1} B_\mu \partial_\nu S -  S^{-1}(\partial_\mu S)
S^{-1}B_\nu S\end{equation}

\noindent
All but the first term (which represents the transformation of $i\epsilon B_\mu
B_\nu$) cancel components of the extraneous terms in equation (4). And
$i\epsilon (B'_\mu B'_\nu - B'_\nu B'_\mu)$ cancels all of the extraneous terms
except the transformation of $i\epsilon(B_\mu B_\nu - B_\nu B_\mu)$.

\noindent
After performing the cancellation, we get

\begin{center}
$ \partial_{\nu}B'_{\mu} -\partial_{\mu}B'_{\nu} +i\epsilon(B'_\mu B'_\nu -
B'_\nu B'_\mu)
 = $
\end{center}
\begin{equation}
S^{-1}[\partial_{\nu}B_{\mu} -\partial_{\mu}B_{\nu} +i\epsilon(B_\mu B_\nu
- B_\nu B_\mu)]S\end{equation}

\noindent
which satisfies equation (1).

\newpage
\epsfig{file =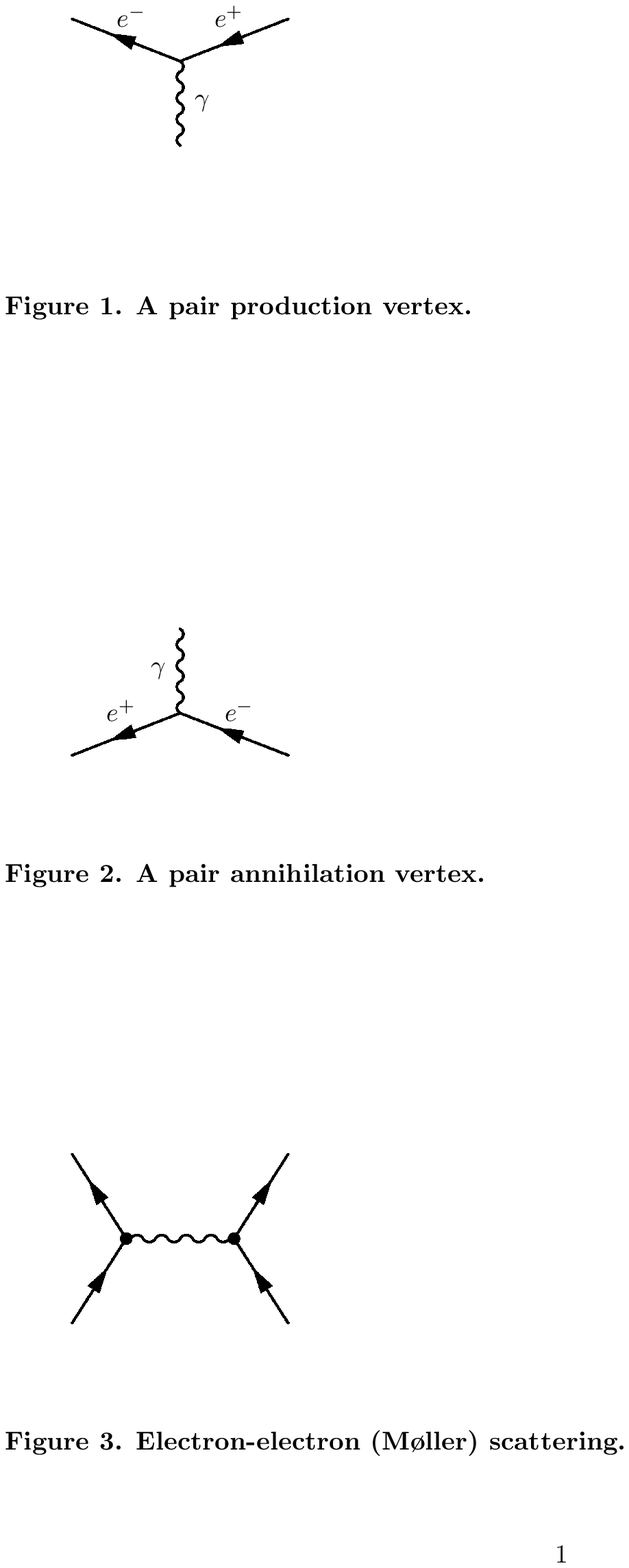}
\newpage
\epsfig{file =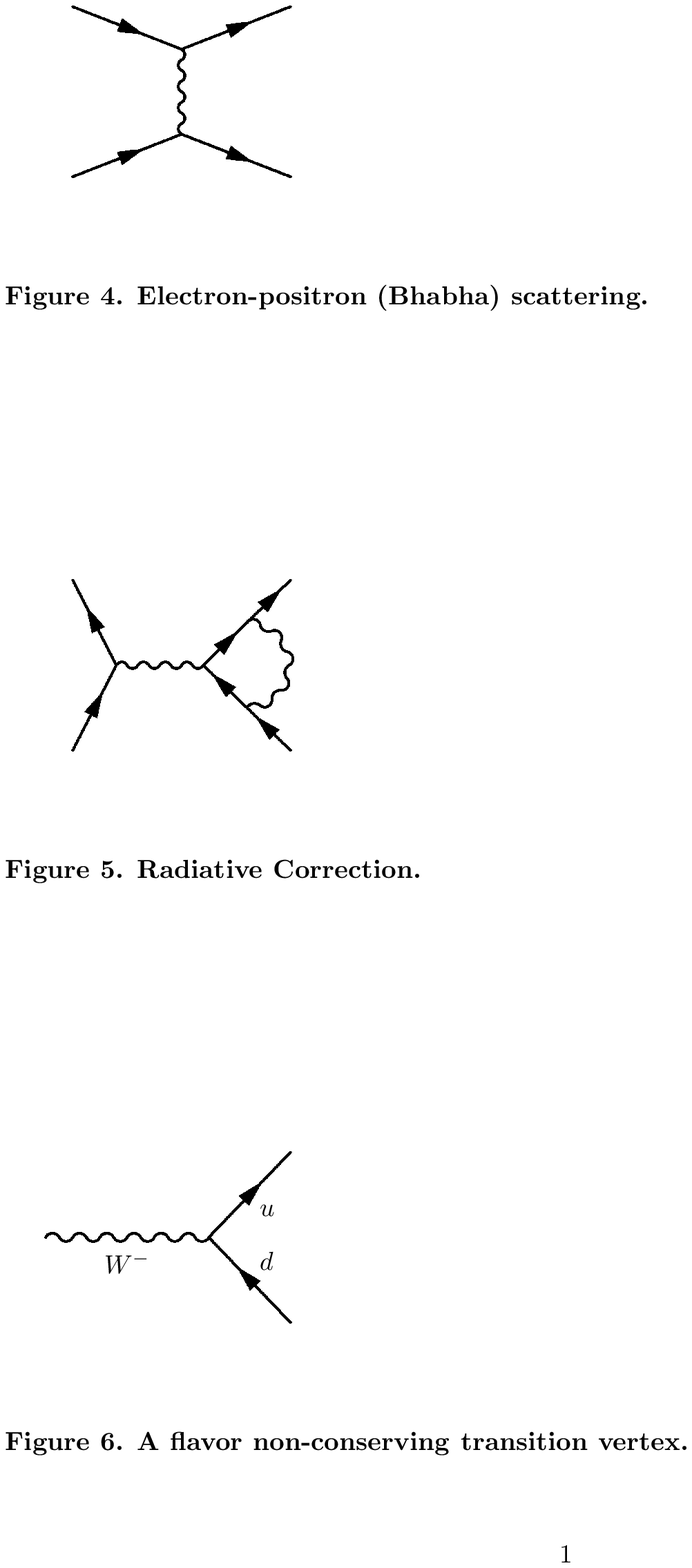}
\end{document}